\documentclass[reprint,amsmath,amssymb,aps]{revtex4}
\usepackage{graphicx}
\begin{document}

%\markboth{Authors' Names}
%{Instructions for Typing Manuscripts (Paper's Title)}

%%%%%%%%%%%%%%%%%%%%% Publisher's Area please ignore %%%%%%%%%%%%%%
%\catchline{}{}{}{}{}
%%%%%%%%%%%%%%%%%%%%%%%%%%%%%%%%%%%%%%%%%%%%%%%%%%%%%%%%%%%%%%%%%%%

\title{Radiation Reaction in Non-commutative Electrodynamics
}

\author{V. Veera Reddy}

\affiliation{BITS Pilani Hyderabad Campus,Hyderabad 500078, India}

\author{Sashideep Gutti}

\affiliation{BITS Pilani Hyderabad Campus,Hyderabad 500078, India
}

\author{Asrarul Haque}

\affiliation{BITS Pilani Hyderabad Campus,Hyderabad 500078, India\\
ahaque@hyderabad.bits-pilani.ac.in}

\date{\today}

%\pub{Received (Day Month Year)}{Revised (Day Month Year)}

\begin{abstract}
We study the radiation reaction acting on an accelerating charge moving in noncommutative spacetime and obtain an expression for it. Radiation reaction, due to a nonrelativistic point charge, is found to receive a \textit{small} noncommutative correction term. The Abraham-Lorentz equation for a point charge in noncommutative spacetime suffers from the preacceleration and the runaway problems. We explore as an application the radiation reaction experienced by a charge which undergoes harmonic oscillations in a noncommutative plane.
%\keywords{Keyword1; keyword2; keyword3.}
\end{abstract}

%\ccode{PACS Nos.: include PACS Nos.}
\maketitle

\section{Introduction}
An accelerated charge emits electromagnetic radiation which carries away certain amounts of the energy and momentum of the charge\cite{jac,swi,bar}. The loss of energy leads to a deceleration of the charge which implies that there must be a force acting on the charge due to the electromagnetic fields it produces. This retarding force is known as the radiation reaction. Thus, the effect caused by the radiation on the dynamics of the charge is to modify the equation of motion of the charge with the radiation reaction.\\
An accelerated charge produces electromagnetic field, say $A^{\mu}(x,y,z,t),$ whose spacetime coordinates commutate among themselves, i.e.,
\[ [x_i,x_j]=0,~~ ~~[x_i,t]=0~;~~i,j=1,2,3.\]
However, there are physical instances where the notion of spatial coordinates may not commute 
\[ [x_i,x_j]\ne ~0~~ ~~~;~~i,j=1,2,3. \]
could be realized\cite{jack}.\\
It is natural to ask: ``How the radiation reaction and hence the dynamics of an accelerated charge will change in a spacetime manifold equipped with noncommuting spacetime coordinates?"\\
% and therefore it loses energy \cite{jac,swi,bar}. Larmor power accounts for the %rate of loss of energy for an accelerated point charge.  Loss of energy could be %conceived to be caused by a damping force operating on the charge produced by the %radiation field. This damping force is known as radiation reaction.\\
%Although radiation reaction in the usual Maxwell theory has been studied extensively %in the past \cite{boy,mon,gri,tem1,tem2,roh}, it continues to be the subject-matter %of several recent studies \cite{haq,aas,ste1,ste2} as well.\\
Physics at high energy, might alter the continuum nature of spacetime, serves as one of the main motivations behind the study of the spacetime coordinates that do not commute. One of the variants of the granular structure of spacetime could be thought in terms of the uncertainties in the measurements of the spacetime coordinates which give rise to the idea of the noncommutative spacetime.
\\
It was Snyder \cite{sny} who first introduced Noncommutative space, as a possible cure of ultraviolet divergences which stem from the ill-defined product of the fields at the same space-time point in quantum field theories. Moreover, noncommutative spaces are found to arise in several different contexts.
At short distance, the interplay between quantum theory and gravitation suggests a non-trivial structure of space-time and a noncommutative structure of space-time is a possibility.
In fact, the concept of space-time as a $c^\infty$ manifold may break down to the distance-scales of the order of Plank length scale \cite{dop}.
Noncommutative structure of spacetime naturally appears in string theory\cite{wit}.
We consider fields defined on noncommutative  spacetime manifold which obeys
\begin{equation}
[\hat x^{\mu},\hat x^{\nu}]=i\theta^{\mu\nu} 
\end{equation}
where $\theta^{\mu\nu}$ is a constant real antisymmetric matrix of length dimension two. Elctromagntic field theory in commutaive space time could be generalized to a noncommutative space-time via replacing ordinary local product by a Moyal star product of the two functions \cite{sza}
\begin{equation}
(f*g)(x):= e^{\frac{i}{2}\theta^{\mu\nu} \partial^x_{\mu}\partial^y_{\nu}}f(x)g(y)|_{x=y} 
\end{equation}
The question arises: can such spacetime noncommutativity induce any modification to the radiation reaction associated with an accelerated charge in commutative spacetime?\\
%According to the classical electrodynamics in commutative spacetime, an accelerating charge radiates and therefore it loses energy \cite{jac,swi,bar}. Larmor power accounts for the rate of loss of energy for an accelerated point charge.  Loss of energy could be conceived to be caused by a damping force operating on the charge produced by the radiation field. This damping force is known as radiation reaction.\\
%Although radiation reaction in the usual Maxwell theory has been studied extensively in the past \cite{boy,mon,gri,tem1,tem2,roh}, it continues to be the subject-matter of several recent studies \cite{haq,aas,ste1,ste2} as well.\\
The goal of the present work is to compute the radiation reaction force experienced by an accelerated charge moving in the noncommutative spacetime. 
%The current work is meant to explore the possible modification present in the radiation reaction force experienced by an accelerating charge in the noncommutative electrodynmics.
We perform the calculation of radiation reaction on an accelerating charge to capture any effect whatsoever arising due to noncommutativity in spacetime with a hope to find the resolutions to the pathologies and ambiguities associated with the Abraham-Lorentz equation. We find that radiation reaction receives modification in the noncommutative Maxwell theory. The Abraham-Lorentz equation in noncommutative spacetime is found to be plagued with the preacceleration and the runaway solutions. This work serves as a preliminary framework to continue further study of the radiation reaction in the quantum version of noncommutative electrodynamics.\\
%In this work, we shall attempt to look at the problem of radiation reaction using Lorentz force density in noncommutative spacetime. 
In section 2, we define and formulate the radiation reaction force in noncommutative spacetime.  In section 3, we derive explicitly the expression for radiation reaction in noncommutative electrodynamics. In section 4, we study radiation reaction experienced by a charge which undergoes harmonic oscillations in a noncommutative plane.
\section{Lorentz force in noncommutative spacetime}
A charged particle moving in an electromagnetic field experiences Lorentz force.
Lorentz four force \cite{bar,jac} on a charged particle in electrodynamics in commutative spacetime is defined as
\begin{equation}
G^{\nu}_{C}=\int k^{\nu} d^3 {x}
\end{equation}
with Lorentz four force density $k^{\nu}$ given by
\begin{equation}
k^{\nu} = \frac{1}{c} j_{\mu} F^{\mu \nu}
\end{equation}
where $ j_{\mu} $ is four current density vector and $F^{\mu \nu} $ is electromagnetic field tensor constructed using the electromagnetic four vector potential $ A^{\nu}$ and is given by $F^{\mu \nu} =(\partial ^{\mu} A^{\nu} - \partial ^{\nu} A^{\mu})$. \\
For a point charge the four current density reads  
\begin{equation}
j^\mu (x)=ec \int_{-\infty}^{+\infty} ds \hspace{2pt} \dot{z}^\mu \delta \left[(x-z(s))\right]
\end{equation}
where $z(s)$ is the world line of the charged particle parametrized by $s$. In order to calculate self force due to the point charge, electromagnetic field tensor $F^{\mu \nu}$ is replaced by retarded field tensor $F^{\mu \nu}_{ret}$ and four potential $A^{\mu}$ is replaced by retarded  potential $A^{\mu}_{ret}$ \cite{bar,roh}.

\begin{align}
F^{\mu \nu}_{ret}=\frac{1}{2} F^{\mu \nu}_{rad}+ \frac{1}{2}  (F^{\mu \nu}_{ret}+F^{\mu \nu}_{adv})
\end{align}
where the radiation field defined by 
\begin{align}
F^{\mu \nu}_{rad}=F^{\mu \nu}_{ret}-F^{\mu \nu}_{adv}
\end{align}
Radiation field $\frac{1}{2} F^{\mu \nu}_{rad}$ contributes to radiation reaction where as $\frac{1}{2}  (F^{\mu \nu}_{ret}+F^{\mu \nu}_{adv})$ contributes to the electromagnetic mass of point charge.\\
Therefore the radiation reaction acting on a point charge in commutative electrodynamics is defined by
\begin{eqnarray*}
G^{\nu}_C &=& \frac{1}{2c}\int  j_{\mu} (x) F^{\mu \nu}_{rad}(x) d^3 {x}. \nonumber 
\end{eqnarray*}
Radiation reaction force \cite{bar} evaluates to
\begin{eqnarray}
G_{iC} &=&-\frac{2}{3} \frac{e^2 \gamma^2}{4 \pi c^3} \left(\ddot{v}_i
+\frac{3 \gamma ^2}{c^2} \dot{v}_i (\vec{v}.\dot{\vec{v}}) + \frac{\gamma ^2}{c^2} v_i (\vec{v}. \ddot{\vec{v}})\nonumber \right.\\
&& \left.+\frac{\gamma ^4}{c^4} (\vec{v}.\dot{\vec{v}})^2 v_i\right)
\end{eqnarray}
where $\textit{i}=1,2,3$ denote the spatial indices and $v_i$ is the three velocity of the charged particle.\\
In the non-relativistic limit, radiation reaction force due to a point charge is given by
\begin{equation}
G^{rad}_{iC}=-\frac{2}{3} \frac{e^2}{4 \pi c^3} \ddot{v}_{i}.
\end{equation}
Lorentz force in commutative electrodynamics could be generalized to noncommutative electrodynamics as follows:
\begin{eqnarray}
F^\nu_{NC}=\int K^\nu\, d^3 x
\end{eqnarray}
with noncommutative Lorentz force density $K^\nu$ \cite{rab} given by 
\begin{eqnarray}
 K^{\nu}&=& \frac{1}{c}\left( j_\mu F^{\mu \nu}+\theta^{\alpha \beta} \partial_{\alpha} \left[ A_{\beta} j_\mu F^{\mu \nu}\right]\right)
\end{eqnarray}
where $A_\mu$ and $j_\mu$ \cite{ado} are given as:
\begin{eqnarray}
A_\mu &=& A_\mu ^{(0)} + A_\mu ^{(1)} (\theta)+ O(\theta ^2)\\
j_\mu &=& j_\mu ^{(0)} + j_\mu ^{(1)} (\theta)+ O(\theta ^2)
\end{eqnarray}
%$j_{\mu},A_{\beta}$and $F^{\mu \nu}$  have the standard interpretation as the commutative electrodynamics.\\
Radiation reaction four force on a charge in noncommutative electromagnetic theory can therefore be defined by
\begin{equation}
G^\nu_{NC}=\int K^\nu_{rad}d^3 x
\end{equation}
where,
\begin{equation}
K_{rad}^{\nu} = \frac{1}{2c}j_\mu  F^{\mu \nu}_{rad}+ \frac{1}{4c} \theta^{\alpha \beta} \partial_{\alpha} \left[ j_\mu A_{\beta}^{rad}  F^{\mu \nu}_{rad}\right]. \label{eq:11}
\end{equation}
%{\bf{
%Equation of motion in NC electrodynamics which is non linear in $f^{\mu \nu}$ is derived  in \cite{ad}.
%That is
%\begin{eqnarray}
%\partial _\nu f^{\nu \mu} -\theta ^{\alpha \beta} \left(\partial _\nu(f^\nu _\alpha f^\mu _\beta) -f_{\nu \alpha} \partial _%%%%\beta f^{\nu \mu} \right)=\frac{4\pi}{c} j^\mu \\
%\textit{ where $j_\mu$ satisfies} \nonumber \\
%\partial_\mu j^\mu + \theta ^{\alpha \beta} f_{\mu \alpha} \partial _\beta j^\mu=0
%\end{eqnarray}
%To treat the nonlinear equations $A_\mu$ and $j_\mu$ are expanded in the $\theta$ - series :
%\begin{eqnarray}
%A_\mu &=& A_\mu ^{(0)} + A_\mu ^{(1)} (\theta)+ O(\theta ^2)\\
%j_\mu &=& j_\mu ^{(0)} + j_\mu ^{(1)} (\theta)+ O(\theta ^2)
%\end{eqnarray}
%Therefore the term  $j_{\mu} F^{\mu \nu}_{rad} $ has $\theta $ dependence. }}
\section{Radiation reaction in noncommutative spacetime}
We shall now turn up to the calculation of the radiation reaction in non-commutative space-time. The radiation reaction in noncommutative spacetime reads:
\begin{eqnarray}
G^{\nu}_{NC} &=&\frac{1}{2c} \int J_{\mu} F^{\mu \nu}_{rad}\, d^3 x \nonumber \\
&+& \frac{\theta^{0 i}}{4c}\int \partial _{0} (A_{i}^{rad} J_\mu F^{\mu \nu}_{rad}) \, d^3 x\nonumber \\
&+& \frac{\theta^{i j}}{4c}\int \partial _{i} (A_{j}^{rad} J_\mu F^{\mu \nu}_{rad}) \, d^3 x 
\end{eqnarray}
where \textit{i, j}=1, 2, 3 are spatial indices and index $0$ stands for the temporal index. All terms containing $\theta^{ij}$ involve total spatial derivative and hence must vanish since the vector potential as well as the electromagnetic field tensor vanish at the large spatial boundary.
The non-trivial surviving terms are given by
\begin{align}
G^{\nu}_{NC}=\frac{1}{2c} \int J_{\mu} F^{\mu \nu}_{rad}\, d^3 x 
+  \frac{\theta^{0 i}}{4c}\int \partial _{0} (A_{i}^{rad} J_\mu F^{\mu \nu}_{rad}) \, d^3 x \label{qq}
\end{align}
The term $\frac{1}{2c} \int J_{\mu} F^{\mu \nu}_{rad}\, d^3 x$  turns out \cite{bar}:
\begin{eqnarray}
\frac{1}{2c}\int J_{\mu} F^{\mu \nu}_{rad} d^3 x &=& \frac{e}{2} \int F^{\mu \nu}_{rad} d^3 x \int \dot{z}_{\mu}(s) \delta(x(t)-z(s)) ds \nonumber \\
%&=&\frac{e}{2} \int \dot{z}_{\mu} \delta(x^0 - z^0(s))F^{\mu\nu}_{rad} (x^0,z(s)) ds \nonumber  \\
&=&\frac{e}{2} \frac{\dot{z}_\mu (s)}{\dot{z}^0 (s)} F^{\mu \nu}_{rad}(z(s))\nonumber \\
&=&-\frac{2}{3} \frac{e^2}{4 \pi } \frac{1}{\dot{z}^0} \left(\dddot{z}^\nu + \dot{z}^\nu \ddot {z}^2  \right)
\label{eq:15}
\end{eqnarray}
The second term to the RHS of the equation \eqref{qq} could be simplified decomposing into two terms (please see Appendix A) as follows:
\begin{eqnarray}
&& \int \frac{\partial}{\partial x^0} (A_i ^{rad}  F^{\mu \nu}_{rad}) J_\mu \, d^3 x  \nonumber \\
&&=\frac{2}{3} \frac{e^3 c}{4 \pi ^2} \left(\dddot{z}^\nu (s) +\dot{z}^\nu (s){\ddot {z}^2(s) } \right) \times \nonumber \\
&& \frac{1}{\dot{z}^0 (s)} \left(\frac{1}{3}\dot{z}_i (s) \dddot {z}_0 (s) + \ddot{z}_i (s) \ddot {z}_0 (s)+\dddot{z}_i (s) \dot {z}_0 (s) \right) \nonumber \\
&& + \frac{e^3 c}{4 \pi ^2} \frac{\ddot{z}_i (s)}{\dot{z}^0 (s)} \left[\frac{ \dddot{z}^0 (s) \ddot{z}^\nu (s)}{6} + \frac{2}{3}\ddot{z}^0 (s) \left(  \dddot{z}^\nu (s)  + \ddot{z} ^2 (s) \dot{z}^\nu (s) \right) \right. \nonumber \\
&& \left. +  \ddot{z} ^2 (s) \ddot{z}^\nu (s)  \left(\frac{1}{2} \dot{z}^0 (s)+ \frac{2}{3} \frac{1}{\dot{z}^0 (s)} \right)\right] \label{fin1}
\end{eqnarray}
and
\begin{eqnarray}
&&\int \frac{\partial J_\mu}{\partial x^0} (A_i ^{rad}  F^{\mu \nu}_{rad})\,  d^3 x \nonumber \\
&&=\frac{2}{3} \frac{e^3 c}{4 {\pi}^2} \frac{\ddot{z}_i (s)}{(\dot{z}^0 (s))^2} \dot{z}^\nu (s) \ddot{z}_\mu (s) \dddot{z}^\mu (s) \nonumber \\
&&-\frac{2}{3} \frac{e^3 c}{4{\pi}^2} \frac{\ddot{z}_i (s) \ddot{z}^0 (s)}{(\dot{z}^0 (s))^3} \left(\dddot{z}^\nu (s) + \ddot{z}^2 (s) \dot{z}^\nu (s) \right)
 \label{fin2}
\end{eqnarray}
So that the radiation reaction becomes
\begin{eqnarray}
&&G^{\nu}_{NC}= -\frac{2}{3} \frac{e^2}{4 \pi } \frac{1}{\dot{z}^0 (s)} \left(\dddot{z}^\nu (s) + \dot{z}^\nu (s) \ddot {z}^2 (s)  \right)\nonumber\\
&&+{\theta}^{0i} \frac{e^3}{16{\pi}^2} \times \nonumber \\
&&\left\lbrace \frac{\ddot{z}^i (s)}{\dot{z}^0 (s)}  \left[\frac{2}{3}\ddot{z}^0 (s) \left(\dddot{z}^\nu (s) + \ddot{z}^2 (s) \dot{z}^\nu (s) \right)\left(\frac{1}{(\dot{z}^0 (s))^2}-2\right) \nonumber \right. \right. \\
&& \left. \left. -  \frac{1}{6} \dddot{z}^0 (s) \ddot{z}^\nu (s) -  \ddot{z} ^2 (s) \ddot{z}^\nu (s)  \left(\frac{1}{2} \dot{z}^0 (s)+ \frac{2}{3} \frac{1}{\dot{z}^0 (s)} \right) \nonumber \right. \right. \\
&& \left. \left. - \frac{\dot{z}^\nu (s) \ddot{z}_\mu (s) \dddot{z}^\mu (s)}{\dot{z}^0 (s)}  \right]  - \frac{2}{3} \left(\dddot{z}^\nu (s) +\dot{z}^\nu (s){\ddot {z}^2(s) } \right) \times \right.  \nonumber \\
&& \left. \frac{1}{\dot{z}^0 (s)} \left(\frac{1}{3}\dot{z}^i (s) \dddot {z}_0 (s) +\dddot{z}^i (s) \dot {z}_0 (s) \right)\right\rbrace \nonumber \\
\end{eqnarray}
Moreover, in terms of 3-velocity, radiation reaction takes the following form
\begin{eqnarray}
G_{j}^{NC}& =&-\frac{2}{3} \frac{e^2 \gamma^2}{4 \pi c^3} \left(\ddot{v}_j +\frac{3 \gamma ^2}{c^2} \dot{v}_j \vec{v}.\dot{\vec{v}} + \frac{\gamma ^2}{c^2} v_j \vec{v}. \ddot{\vec{v}} +\frac{3 \gamma ^4}{c^4} (\vec{v}.\dot{\vec{v}})^2 v_j   \right) \nonumber \\
& -& \theta ^{0i} \frac{e^3 \gamma ^6}{16 \pi ^2 c^6} \left(\frac{2}{3}\ddot{v}_i \ddot{v}_j - \dot{v}_i v_j \dot{\vec{v}}.\ddot{\vec{v}}\right) 
\end{eqnarray}
where $\textit{i,~j}=1,2,3$.\\
Thus the radiation reaction force is modified by $\theta-$dependent terms and it, in fact, depends on both space-space $\theta^{ij}$ and space-time $\theta^{0i}$ non commutative parameters due to $\theta ^{\mu \nu} $ dependence of velocity $v_j$ and its various order time derivatives.\\
In the non-relativistic limit $v/c\to 0$, radiation reaction force turns out to be
\begin{eqnarray}
G_ {j}^{NC} &=&-\frac{2}{3} \frac{e^2 }{4 \pi c^3} \ddot{v}_j \left(1 + \theta ^{0i}  \frac{e}{4 \pi c^3}\ddot{v}_i \right). \label{eqrr}
\end{eqnarray}
The non-relativistic expression for radiation reaction in noncommutative electrodynamics, unlike its commutative counterpart, receives $\theta$-correction term. Even in the case of zero space-time noncommutativity $\theta ^{0i}=0$,  
% owing to its dependence on non-commutative parameter $\theta^{\mu\nu}$ through $\ddot{v}_j (\theta )$.
space-space noncommutativity $\theta^{ij}\ne 0 $ is there through $\theta^{\mu\nu}$ dependence of $\ddot{v}_j$ which could be seen in the example of radiation reaction experienced by charged harmonic oscillator in noncommutative plane which is discussed in the Section 4.
It is worthy to note at this juncture that, the non relativistic limit  $ \frac{v}{c} \rightarrow 0 $ doesn't imply the commutative limit $\theta^{\mu\nu} \rightarrow 0$ and vice versa. 
The Abraham-Lorentz equation in noncommutative spacetime now reads:
\begin{equation} ma_j - m\tau \dot{a_j}- \frac{3m^2 \tau ^2 }{2e} \theta^{0i}\dot{a_i} \dot{a_j} =F_j \end{equation}
where $\dot{v}_j=a_j,$ the characteristic time $\tau = \frac{2}{3} \frac{e^2 }{4 \pi m c^3}$ and $F_j$ is the external force.\\
Let us consider one space - one time dimensions and assume $\theta ^{01}=\theta $ then  the radiation reaction force becomes 
\begin{eqnarray}
G^{NC} &=&-\frac{2}{3} \frac{e^2 }{4 \pi c^3}  \left(\ddot{v} + \theta   \frac{e}{4 \pi c^3}\ddot{v} ^2 \right). \label{NCf}
\end{eqnarray}
%Characteristic time $\tau$ is defined as 
%\begin{eqnarray}
%&& \tau = \frac{2}{3} \frac{e^2 }{4 \pi m c^3}
%\end{eqnarray}
Now, the equation of motion of a charge particle under a constant external force $F$ is given by
\begin{equation}
\frac{3}{2} \frac{m \tau ^2 \theta}{e} \dot{a} ^2 + \tau \dot{a} -a+\frac{F}{m} =0\label{eqm}
\end{equation}
Equation \eqref{eqm} is quadratic in $\dot{a}$. The acceptable root which is consistent in the commutative limit $\theta \to 0$ is

\begin{equation}
\dot{a} = \frac{e}{3m \tau  \theta} \left(-1 +  \sqrt{1 + \frac{6 m \theta}{e}(a-F/m)}\right) \label{s1}
\end{equation}
Suppose the constant external force $F$ acts on the charge only during the time interval $0 \leq t \leq T$. Therefore,
in the region $t \leq 0,$   $F = 0$ and equation \eqref{s1} leads to \\
\begin{equation}
\dot{a} = \frac{e}{3m \tau  \theta} \left(-1 +  \sqrt{1 + \frac{6 m \theta}{e}}a\right) 
\end{equation} 
which, after integration, yields
\begin{eqnarray}
&&(-1+ \sqrt{1+\frac{6m\theta a_1}{e} }) \exp (-1+ \sqrt{1+\frac{6m\theta a_1}{e} })=I_1 \exp{\frac{t}{\tau}}  
\label{a3}
\end{eqnarray}

where $I_1$ is the integration constant.
In the region $0 \leq t \leq T,$ we have
\begin{equation}
(-1+ \sqrt{1+\frac{6m\theta(a_2-\frac{F}{m})}{e} }) \exp (-1+ \sqrt{1+\frac{6m\theta (a_2-\frac{F}{m})}{e} }) 
=I_2 \exp{\frac{t}{\tau}}
\end{equation}
 
Similarly for the region $ t\geq T $, external force F is zero and the connection between acceleration and time in this region turns out 
\begin{eqnarray}
&&(-1+ \sqrt{1+\frac{6m\theta a_3}{e} }) \exp (-1+ \sqrt{1+\frac{6m\theta a_3}{e} })=I_3\exp{\frac{t}{\tau}}  
\label{a4}
\end{eqnarray}
$I_2$ and  $I_3 $ are integration constants. The acceleration versus time in the region $t\geq T$ with $T=1\tau$ is plotted as shown in the Figure 1.

\begin{figure}[ht]
\centering
\includegraphics[scale=0.7]{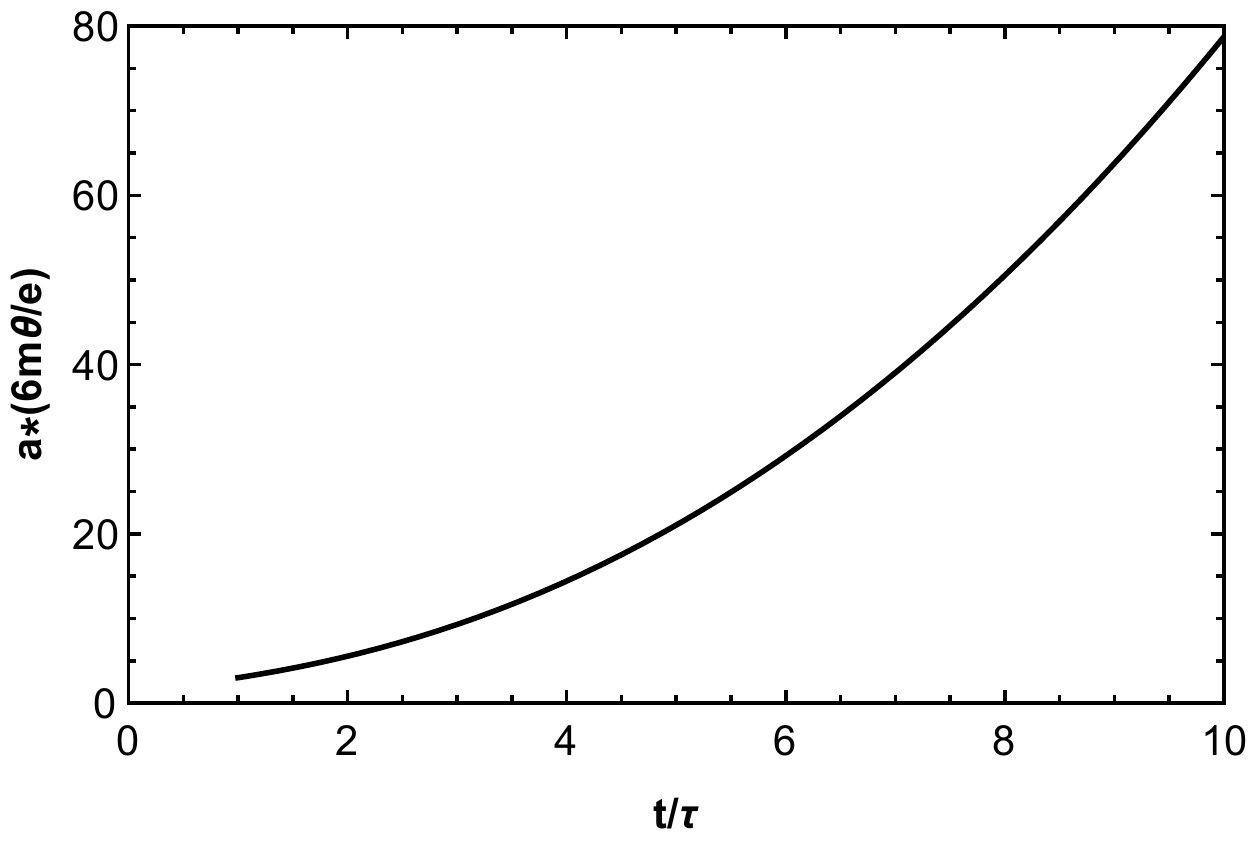}
\caption{}
\end{figure}
We notice that the acceleration in the region $t\geq T$ is increasing continuously with time $t$.  Therefore, equation \eqref{a4} will admit runaway solutions.\\
Thus, Abraham-Lorentz equation in noncommutative spacetime is plagued with preacceleration and runaway solutions.
\section{Radiation reaction in two dimensional noncommutative harmonic oscillator}
In this section, we explore the radiation reaction force due to a charge undergoing simple harmonic motion in a noncommutative plane $([\hat x,\hat y]=i\theta)$.\\ 
The nonrelativistic Lagrangian for 2-dimensional noncommutative harmonic oscillator \cite{dra} in terms of commutative coordinates $q_1$ and $q_2$ to the first order in $\theta$ is given by: 
%$$L=\frac{m}{2 \kappa} \left(\dot{q}_{1}^2 + \dot{q}_{2}^2\right)+\frac{m^2 \omega^2 \theta}{2 \kappa} \left(\dot{q}_2 q_1 - %%%\dot{q}_1 q_2 \right)-\frac{m \omega ^2}{2 \kappa} \left({q}_{1}^2 + {q}_{2}^2 \right)$$
%where $\kappa=1+\frac{m^2 \omega^2 \theta ^2}{4}$ and to the first order in $\theta$ , $\kappa=1$. Hence, Lagrangian to %%%$O(\theta)$ equals 
\begin{equation}
L=\frac{m}{2 } \left(\dot{q}_{1}^2 + \dot{q}_{2}^2\right)+\frac{m^2 \omega^2 \theta}{2 } \left(\dot{q}_2 q_1 - \dot{q}_1 q_2 \right)-\frac{m \omega ^2}{2 } \left({q}_{1}^2 + {q}_{2}^2 \right).
\end{equation}
The Euler-Lagrange equations in $q_1$ and $q_2$, unlike commutative counterpart, are coupled equations as given below:
\begin{eqnarray*}
\ddot{q}_1 - m \omega^2 \theta \dot{q}_2 + \omega^2 q_1 =0  \\
\ddot{q}_2 + m \omega^2 \theta \dot{q}_1 + \omega^2 q_2 =0.  \\
\end{eqnarray*}
%{\bf As can be observed, the classically independent equations have become coupled set of equations. By defining,}
The effect of noncommutativity is to couple the two independent modes and hence the system now oscillates with two distinct normal mode frequencies, say $\omega_+$ and $\omega_-$.
The general solution of the above differential equations are given by
\begin{eqnarray*}
q_1(t)&=& C_1 \cos(\omega_+ t) + C_2 \sin(\omega_+ t)\nonumber \\
&&{}+ C_3 \cos(\omega_- t) + C_4 \sin(\omega_- t)\\ 
q_2(t)&=& D_1 \cos(\omega_+ t) + D_2 \sin(\omega_+ t) \nonumber \\
&&{}+ D_3 \cos(\omega_- t) + D_4 \sin(\omega_- t)
\end{eqnarray*}
where,
\begin{eqnarray*}
\omega _+& =&\omega+\frac  {m \theta \omega^2}{2} \\
\omega _-&=&-\omega+\frac  {m \theta \omega^2}{2}.
\end{eqnarray*}
$C_i$ and $D_i$ are integration constants to be fixed by the initial conditions. Moreover, there are only four independent constants $(C_1=-D_2, C_2=D_1, C_3=-D_4, C_4 =D_3)$ out of the eight ones due to the differential relations between $q_1$ and $q_2$. The radiation reaction experienced by the charged harmonic oscillator in the noncommutative plane using \eqref{eqrr} is given by
\begin{eqnarray}
\vec{F}_{rad}&=&-\frac{2}{3} \frac{e^2}{4 \pi c^3}[{ \omega _ + }^3 (C_1 \sin { \omega _ + t} -  C_2 \cos{ \omega _ + t}) \nonumber\\
&+&{\omega _ -}^3 ( C_3 \sin { \omega _ - t} -  C_4 \cos{ \omega _ - t})] \hat{q}_1\nonumber\\
&-&\frac{2}{3} \frac{e^2}{4 \pi c^3}[{\omega _ + }^3 (C_1 \cos { \omega _ + t} +  C_2 \sin {\omega _ + t}) \nonumber\\
&+&{\omega _ -}^3 ( C_3 \cos { \omega _ - t} +  C_4 \sin { \omega _ - t})] \hat{q}_2
\end{eqnarray}
Where $\hat{q}_1$ and $\hat{q}_2$ are unit vectors along $q_1$ and $q_2$ directions respectively. We shall commit to the specific case when the initial conditions are such that $C_1=C_2=C_3=C_4 =1$. This choice does not alter the generalities of the interpretation of result for radiation reaction force due to 2D noncommutative charged harmonic oscillator. The quantity $ \frac {m \theta \omega^2 t}{2}$ for all practical purposes is rather small and could admit large enough time to observe radiation reaction. Therefore, we can have for our purpose $\sin\frac {m \theta \omega^2 t}{2}\sim \frac {m \theta \omega^2 t}{2}$, so that 
\begin{eqnarray*}
\cos\left(\pm \omega+\frac  {m \theta \omega^2}{2}\right)t &\sim&\cos{\omega t}
\mp \left(\frac{m \theta \omega^2 t}{2} \right) \sin{\omega t}\\
\sin\left(\pm \omega+\frac  {m \theta \omega^2}{2}\right)t &\sim&\pm \sin{\omega t} 
+ \left(\frac{m \theta \omega^2 t}{2} \right)\cos{\omega t}\\
%\cos\left(-\omega+\frac  {m \theta \omega^2}{2}\right)t &=&\cos{\omega t}+ \left(\frac{m \theta \omega^2 t}{2} \right) %%\sin{\omega t}\\
%\sin\left(-\omega+\frac  {m \theta \omega^2}{2}\right)t &=&-\sin{\omega t} + \left(\frac{m \theta \omega^2 t}{2} \right)%%%%%%%%\cos{\omega t}
\end{eqnarray*}
so that,
 \begin{eqnarray}
q_1&=&2 \cos{\omega t}+ \left(m \theta \omega^2 t \right)\cos{\omega t} \nonumber\\
q_2&=&2 \cos{\omega t}- \left(m \theta \omega^2 t \right)\cos{\omega t}\nonumber \\
\dddot{q}_1&=&2 \omega^3 \sin \omega t - m \theta \omega^4 (3  \cos \omega t-\omega t \sin \omega t) \label{eq:25}\\
\dddot{q}_2&=&2 \omega^3 \sin \omega t + m \theta \omega^4 (3  \cos \omega t  - \omega t \sin \omega t)\label{eq:26}
\end{eqnarray}
Radiation reaction turns out
%\begin{equation}
%\vec{F}_{Rad}=\hat{q_1}F^1_{Rad}+\hat{q_2}F^2_{Rad}
%\end{equation}
%where,
%\begin{eqnarray*}
%F^1_{Rad}&=&-\frac{2}{3} \hspace{2pt} \frac{e^2}{4 \pi c^3} \left(2 \omega^3 \sin \omega t -  m \theta \omega^4 (3  \cos \omega %t -\omega t \sin \omega t) \right)
%\\
%F^2_{rad}&=&-\frac{2}{3} \hspace{2pt} \frac{e^2}{4 \pi c^3} \left(2 \omega^3 \sin \omega t +  m \theta \omega^4 (3  \cos \omega %t -\omega t \sin \omega t)\right)
%\end{eqnarray*}
%We can rewrite the radiation reaction as sum of $\theta$ independent and $\theta$ dependent pieces as follows:
\begin{equation}
\vec{F}_{Rad}=(\hat{q_1} + \hat{q_2})f(t)+(\hat{q_1} - \hat{q_2}) \theta g(t)
\end{equation}
where 
\begin{eqnarray}
f(t)&=&-\frac{2}{3} \hspace{2pt} \frac{e^2}{4 \pi c^3} 2 \omega^3 \sin \omega t \\
g(t)&=&\frac{2}{3} \hspace{2pt} \frac{e^2}{4 \pi c^3} m  \omega^4 (3  \cos \omega t -\omega t \sin \omega t).
\end{eqnarray}
Let us define
\begin{eqnarray}
f_s(t)&=&\frac{f(t)}{-\frac{2}{3} \hspace{2pt} \frac{e^2}{4 \pi c^3} 2 \omega^3} \\
g_s(t)&=& \frac{g(t)}{\frac{2}{3} \hspace{2pt} \frac{e^2}{4 \pi c^3} m  \omega^4}.
\end{eqnarray}
Figure 2 shows a plot between the scaled components $f_s(t)$ and $g_s(t)$ of the radiation reaction and time. 
%In order to plot the graph, we have used $\omega\sim 10^6 Hz$.
%\begin{center}
\begin{figure}[ht]
\centering
\includegraphics[scale=0.5]{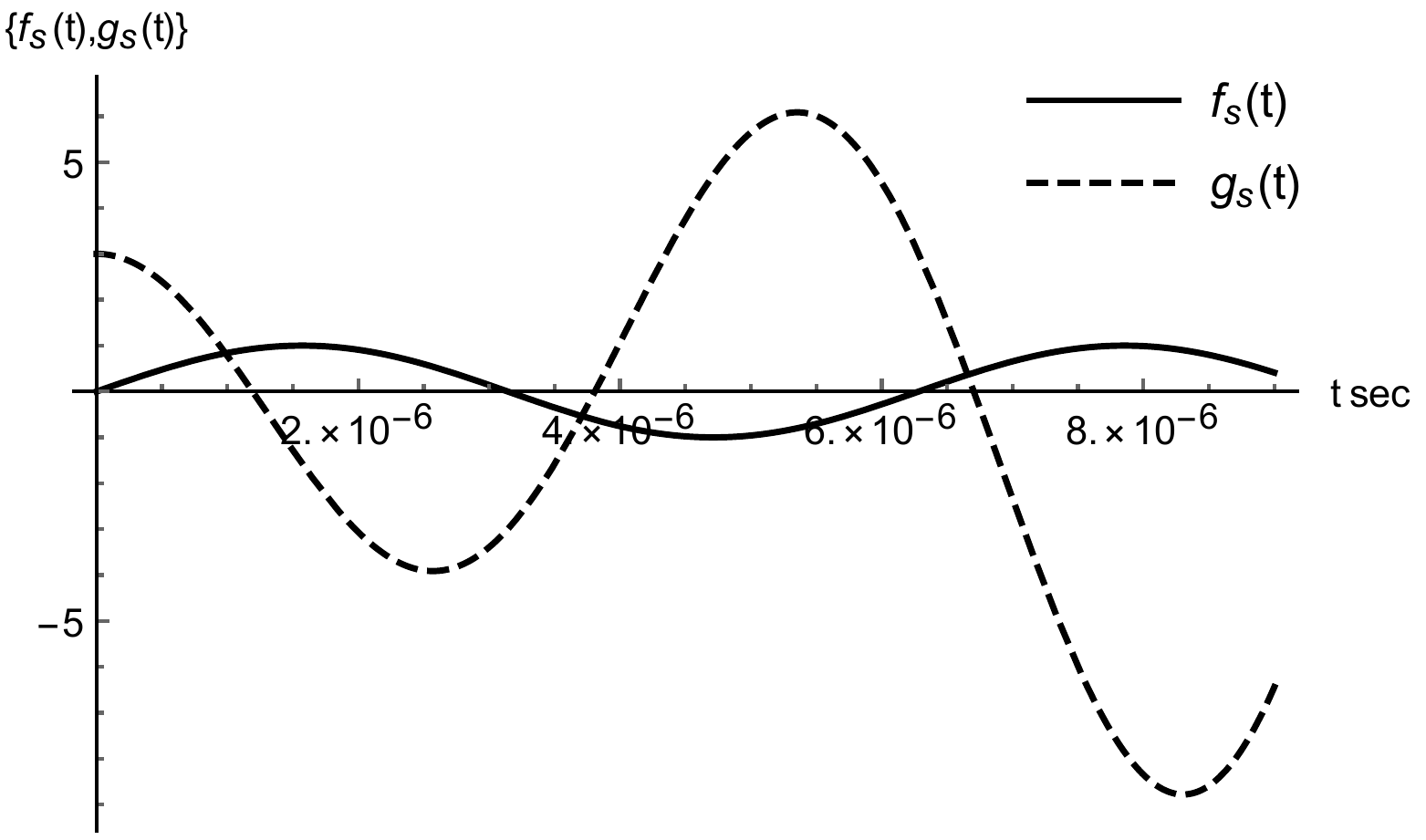}
\caption{}
\end{figure}
%\end{center}
%%%%%%%
Radiation reaction in the commutative space vanishes at all times $t= n\pi/\omega,$ where $n = 0,1,2,3...$. However, radiation reaction in the non-commutative space does not vanish at any point in time.
\section{Conclusion}
The present paper explores whether Maxwell's electromagnetic theory in the noncommutative spacetime might affect the radiation reaction and could possibly resolve the pathologies and ambiguities associated with the Abraham-Lorentz equation. To this end, we have derived the radiation reaction force acting on an accelerating charge due to the electromagnetic fields that it produces while moving in the noncommutative spacetime. We find that an accelerating point charge in noncommutative Maxwell theory experiences radiation reaction which receives a small $\theta-$dependent correction term. Moreover, it turns out that the Araham-Lorentz equation in noncommutative spacetime suffers from the same long-standing problems of the preacceleration and runaway solutions.
%%%%%%%%%%%%%%%%%%%%%%%%%%%%%%%%%%%%%%%%%%%%%%%%%%%%%%%%%%%%%%%%%
\section*{Appendix A : Calculation of radiation reaction in noncommutative electrodynamics}
We have
\begin{eqnarray}
&&\int d^3 x \, \partial _0 \left(A_i ^{rad}(x) J_\mu (x) F^{\mu \nu} _{rad} (x)\right) \nonumber \\
&&=\int d^3 x \, \frac{\partial}{\partial x^0} \left(A_i ^{rad}(x)  F^{\mu \nu}_{rad} (x)\right) J_\mu (x) \nonumber \\
&&+ \int d^3 x \, \frac{\partial J_\mu (x)}{\partial x^0} A_i ^{rad}(x)  F^{\mu \nu}_{rad}(x) \nonumber \\
 \label{ft1} 
\end{eqnarray}
%First term in the above equation is simplified as 
%\begin{eqnarray}
%&&\int d^3 x \, \frac{\partial}{\partial x^0} (A_i ^{rad}(x)  F^{\mu \nu}_{rad} (x)) J_\mu (x) \, =   \nonumber \\
%&&ec \int d^3 x \, \frac{\partial}{\partial x^0}  \left( A_i ^{rad}  F^{\mu \nu}_{rad}\right)  \int ds \, \dot{z}_\mu (s) %\delta(x-z(s))  \nonumber \\
 %\label{1ft1}
%\end{eqnarray}
Four vector current density due to a point charge is given by
\begin{eqnarray}
J_\mu(x)&=& ec  \int ds \, \dot{z}_\mu (s) \delta^4 (x-z(s)) \nonumber \\
%&=&\frac{\dot{z}_\mu(s)\, \delta ^3 (\vec{x}-\vec{z}(s))}{\frac{dz^0 (s)}{ds}} \rvert _{x^0=z^0(s)} \nonumber \\
&=&ec\frac{\dot{z}_\mu (s) \, \delta ^3 (\vec{x}-\vec{z}(s))}{\dot{z}^0 (s)}\rvert _{x^0=z^0(s)} 
 \label{jd}
\end{eqnarray}
%Now the equation \eqref{1ft1} becomes 
The first term in the above equation after inserting \eqref{jd} gets simplified as  
\begin{eqnarray}
&&\int d^3 x \, \frac{\partial}{\partial x^0} (A_i ^{rad}(x)  F^{\mu \nu}_{rad} (x)) J_\mu (x)  \nonumber \\
%&=& ec \frac{\dot{z}_\mu (s)}{\dot{z}^0 (s)} \int d^3 x \, \delta ^3 (\vec{x}-\vec{z}(s)) \left(\frac{\partial}{\partial x^0} (A_i ^{rad}(x)  F^{\mu \nu}_{rad} (x))\right)_{x^0 =z^0 (s)}  \nonumber \\
%&=&ec \frac{\dot{z}_\mu (s)}{\dot{z}^0 (s)}  \left(\frac{\partial}{\partial x^0} (A_i ^{rad} (x)  F^{\mu\nu}_{rad} (x))\right)_{x=z(s)}  \nonumber \\
&=&ec \frac{\dot{z}_\mu (s)}{\dot{z}^0 (s)} F^{\mu \nu}_{rad}(z) \left(\frac{\partial A_i^{rad}}{\partial x^0}\right)_{x=z(s)} \nonumber \\
&+&ec \frac{\dot{z}_\mu (s)}{\dot{z}^0 (s)} A_i^{rad}(z)\left(\frac{\partial F^{\mu \nu}_{rad}}{\partial x^0}\right)_{x=z(s)} \label{ft11}
\end{eqnarray}
The second term in the equation \eqref{ft1} becomes
\begin{eqnarray}
&&{}\int   d^3 x \frac{\partial J_\mu (x)}{\partial x^0} (A_i ^{rad} (x)  F^{\mu \nu}_{rad}(x)) \nonumber\\
%&&{} ec\int d^3 x \, A^{rad}_{i} (x) F^{\mu \nu}_{rad} (x)\frac{\partial}{\partial x^0} \int ds \, \dot{z}_\mu (s) \delta(x-z(s))  \nonumber\\
%&&{}=ec \int d^3 x \,  A^{rad}_i(x)F^{\mu \nu}_{rad}(x)  \int ds \, \dot{z}_\mu (s) \frac{\partial \delta(x-z(s))}{\partial x^0} 
%\end{eqnarray*}
%Using  
%\begin{equation*}
%\frac{\partial \delta (x-z(s)) }{\partial x^0}= -\frac{\partial \delta (x-z(s))}{\partial z^0},
%\end{equation*}
% the above equation becomes \\
%\begin{eqnarray*}
%&=&-ec \int d^3 x \, A^{rad}_i(x)F^{\mu \nu}_{rad}(x)  \int ds \, \dot{z}_\mu (s) \frac{\partial \delta(x-z(s))}{\partial z^0}  \nonumber \\
%&=&-ec  \int d^3 x \, A^{rad}_i(x)F^{\mu \nu}_{rad}(x)  \int  ds\, \frac{\dot{z}_\mu (s)}{\dot{z}^0 (s)} \frac{ \partial \delta(x-z(s))}{\partial s}   \nonumber \\
%\end{eqnarray*}
%\begin{eqnarray*}
%&=&ec \int d^3 x \, A^{rad}_i(x)F^{\mu \nu}_{rad}(x)  \int ds \, \frac{\partial}{\partial s}\left(\frac{\dot{z}_\mu (s)}{\dot{z}^0 (s)}\right) \delta(x-z(s))  \nonumber  \\
%\end{eqnarray*}
%\begin{eqnarray} 
&&=ec \frac{A^{rad}_i(z)}{\dot{z}^0 (s)} {F^{\mu \nu}_{rad}(z)} \frac{\partial}{\partial s} \left(\frac{\dot{z}_\mu (s)}{\dot{z}^0 (s)}\right) \label{ft2}
\end{eqnarray}
Four vector potential is defined as %\cite{barut}
\begin{eqnarray*}
A_\mu ^{rad} (x)%&=&\frac{1}{c}\int  d x^{\prime} \, D(x-x^{\prime}) J_\mu (x^{\prime}) \\
%&=&e\iint  dx^\prime \, ds \, D(x-x^{\prime})\dot{z}_\mu (s) \delta(x^{\prime}-z(s))\\
&=& e\int   ds \, D(x-z(s))\dot{z}_\mu (s)
\end{eqnarray*}
At $x=z(s^{\prime})$, we have
\begin{eqnarray}
A_\mu ^{rad} z(s^{\prime}) &=& e \int D(z(s^{\prime})-z(s))\dot{z}_\mu (s) ds
\end{eqnarray}
Let $ s=s^{\prime} + u $ where u is a small parameter.% \cite{barut}.
\begin{eqnarray*}
z_\mu (s)&=& z_\mu(s^{\prime}) + u \dot{z}_\mu (s^{\prime})+\frac{u^2}{2} \ddot{z}_\mu (s^{\prime})+\frac{u^3}{6} \dddot{z}_\mu  (s^{\prime})+... \\
\dot{z}_\mu (s)&=& \dot{z}_\mu(s^{\prime}) + u \ddot{z}_\mu (s^{\prime})+\frac{u^2}{2} \dddot{z}_\mu (s^{\prime})+...
\end{eqnarray*}
We know \cite{bar}
\begin{eqnarray*}
D(z(s^{\prime})-z(s))&=&\frac{1}{2 \pi} \frac{d \delta (u)}{du} 
\end{eqnarray*}
Therefore
\begin{eqnarray}
A^{rad}_{\mu} (z(s^{\prime}))%= \nonumber \\
%&&\frac{e}{2 \pi} \int \frac{d \delta (u)}{du} du \left( \dot{z}_\mu(s^{\prime}) + u \ddot{z}_\mu (s^{\prime}) +\frac{u^2}{2} \dddot{z}_\mu (s^{\prime})+..\right) \nonumber\\
= -\frac{e}{2 \pi} \int \delta(u)  du \, (\ddot{z}_\mu (s^{\prime})+ u \dddot{z}_\mu (s^\prime)+..)   \nonumber 
\end{eqnarray}
At $u=0 ,s=s^\prime$. Hence
\begin{eqnarray}
A^{rad}_{\mu} (z(s))&=&  -\frac{e}{2 \pi} \ddot{z}_\mu (s)  \label{A}
\end{eqnarray}
%\vspace{0.5pt}
%%%%%%%%%%%%%%%%%%%%%%%%%%%%%%%%%%%%%%%%%
\section*{ {A1: Calculation of}
$\mathbf{(\partial_\nu A_\mu ^{rad})_{x= z(s)}:}$}
\begin{eqnarray*}
&&\partial_\nu A_\mu ^{rad}(x)=e \frac{\partial}{\partial x^\nu} \int ds \, D(x-z(s)) \dot{z}_\mu (s)  \\
%&&=-e \int ds \, \dot{z}_\mu (s) \frac{dD(x-z(s))}{ds} \frac{(x-z(s))_\nu}{(x-z(s))^{\sigma} \dot{z}_{\sigma} (s)} \\
&&= e \int ds \, D(x-z(s))  \frac{d}{ds} \left[  \frac{ \dot{z}_\mu (s)(x-z(s))_\nu}{(x-z(s))^{\sigma} \dot{z}_{\sigma}(s)}\right]  \\  
\end{eqnarray*}
At $x=z(s^\prime)$ 
\begin{eqnarray*}
&&\partial_\nu A_\mu ^{rad}(z(s^\prime))= \nonumber \\
&& e \int ds\, D((z(s^{\prime})-z(s)) 
\frac{d}{ds} \left(\frac{\dot{z}_\mu (s)(z(s^{\prime})-z(s))_\nu}{(z(s^{\prime})-z(s))^{\sigma} \dot{z}_{\sigma}}\right)\nonumber \\ 
\end{eqnarray*}
Now:
\begin{equation}
%&&(z(s^{\prime})-z(s))_\nu =-u \dot{z}_\nu (s^{\prime})-\frac{u^2}{2} \ddot{z}_\nu (s^{\prime})- \frac{u^3}{6}
% \dddot{z}_\nu %(s^{\prime})...\nonumber \\ 
(z(s^{\prime})-z(s))^{\sigma} \dot{z}_{\sigma} = -u+o(u^3) 
\end{equation}
\begin{eqnarray}
&&\dot{z}_\mu (s)(z(s^{\prime})-z(s))_\nu \nonumber \\
&&=-u \left\lbrace \dot{z}_\mu (s^{\prime}) \dot{z}_\nu (s^{\prime})+\frac{u}{2}  \dot{z}_\mu (s^{\prime}) \ddot{z}_\nu (s^{\prime})+ u \ddot{z}_\mu (s^{\prime}) \dot{z}_\nu (s^{\prime})  \nonumber \right. \nonumber \\
&&\left. + \frac{u^2}{6} \dot{z}_\mu (s^{\prime}) \dddot{z}_\nu (s^{\prime}) +\frac{u^2}{2} \ddot{z}_\mu (s^{\prime}) \ddot{z}_\nu (s^{\prime}) \right. \nonumber \\
&& \left. +\frac{u^2}{2} \dddot{z}_\mu (s^{\prime}) \dot{z}_\nu (s^{\prime})+... \right\rbrace \label{eq:36} 
\end{eqnarray}
%%%%%%%%%%%%%%%%%%%%%%%%%%%%%%%%%%%%%%%%%%%%%%%%%%%%%%%%%%%%%%%%%%%%%%%%%%%%%%%%%%%%%%%%%%%%%
%\begin{eqnarray*}
%&&(\partial_\nu A_\mu ^{rad} )_{x=z(s^{\prime})} \\
%&=&-\frac{e}{2\pi} \frac{d^2}{du^2} \left(\dot{z}_\mu (s^{\prime}) \dot{z}_\nu (s^{\prime})+\frac{u}{2}  \dot{z}_\mu %(s^{\prime}) \ddot{z}_\nu (s^{\prime}) \right. \\
%&+&\left.  u \ddot{z}_\mu (s^{\prime}) \dot{z}_\nu (s^{\prime}) +\frac{u^2}{6} \dot{z}_\mu (s^{\prime}) \dddot{z}_\nu %(s^{\prime})\right. \\
%&+&\left. \frac{u^2}{2} \ddot{z}_\mu (s^{\prime}) \ddot{z}_\nu (s^{\prime})+\frac{u^2}{2} \dddot{z}_\mu (s^{\prime}) \dot{z}_\nu %(s^{\prime}) \right)_ {u=0,s=s^{\prime}}
%\end{eqnarray*}
\begin{eqnarray}
&&(\partial_{\nu} A_{\mu} ^{rad})_{x= z(s)}= \nonumber \\
&&-\frac{e}{2\pi}  \left(\frac{1}{3}\dot{z}_\mu (s)\dddot {z}_\nu (s) + \ddot{z}_\mu (s)\ddot {z}_\nu (s)+\dddot{z}_\mu (s)\dot {z}_\nu (s) \right)\nonumber \\
 \label{dA}
\end{eqnarray}
Therefore:
\begin{eqnarray}
F^{\mu \nu}_{rad}(z(s))&=& \left(\partial ^\mu A^\nu _{rad} - \partial ^\nu A^\mu _{rad} \right)_{x=z(s)}\nonumber \\
&=& \frac{e}{3\pi} \left(\dot{z}^\nu (s)\dddot {z}^\mu (s)-\dddot{z}^\nu (s)\dot {z}^\mu (s) \right) \label{Fmu}
\end{eqnarray}
%\vspace{1.0pt}
%%%%%%%%%%%%%%%%%%%%%%%%%%%%%%%%%%%%%%%%
\section*{{A2: Calculation of the term }}
\hspace{0.5in}
$ \mathbf {ec \frac{\dot{z}_\mu (s)}{\dot{z}^0 (s)} F^{\mu \nu}_{rad} (z(s)) (\frac{\partial A_i ^{rad}}{\partial x^0}) _{x= z(s)}:}$\vspace{0.1in}\\ 
%%%%%%%%%%%%%%%%%%%%%%%%%%%%%%%%%%%%%%%
From equation \eqref{dA}
%\vspace{0.3in}
\begin{eqnarray*}
&&(\partial_0  A_i ^{rad})_{x= z(s)} = \nonumber \\
&&-\frac{e}{2\pi} \left(\frac{1}{3}\dot{z}_i (s) \dddot {z}_0 (s)  + \ddot{z}_i (s) \ddot {z}_0 (s) +\dddot{z}_i (s) \dot {z}_0 (s)  \right)
\end{eqnarray*}
Therefore
\begin{eqnarray*}
&&ec \frac{\dot{z}_\mu (s)}{\dot{z}^0 (s)} F^{\mu \nu}_{rad} (z(s)) (\partial_0 A_i ^{rad}) _{x= z(s)} = \nonumber \\
&&- \frac{\dot{z}_\mu (s)}{\dot{z}^0 (s)} \frac{2}{3} \frac{e^2 c}{2\pi} \left(\dot{z}^\nu (s)\dddot {z}^\mu (s)-\dddot{z}^\nu (s)\dot {z}^\mu (s) \right) \times \nonumber \\
&&{} \frac{e}{2\pi} \left(\frac{1}{3}\dot{z}_i (s) \dddot {z}_0 (s)  + \ddot{z}_i (s) \ddot {z}_0 (s) +\dddot{z}_i (s) \dot {z}_0 (s)  \right) 
\end{eqnarray*}
We know that :
\begin{eqnarray*}
&&\dot{z}^\mu (s) \dot{z}_\mu (s) =1, \dot{z}_\mu (s) \ddot{z}^\mu (s) =0 \\
&&\dot{z}_\mu (s) \dddot{z}^\mu (s) +\ddot{z}_\mu (s) \ddot{z}^\mu (s) =0 \\
&&\dot{z}_\mu (s) \dddot{z}^\mu (s) =-{\ddot{z}}^2 (s)
\end{eqnarray*}
Now,
\begin{eqnarray}
&&ec\frac{\dot{z}_\mu (s)}{\dot{z}^0 (s)} F^{\mu \nu}_{rad} (z(s)) (\partial_0 A_i ^{rad}) _{x= z(s)}\nonumber  \\
&& = \frac{2}{3} \frac{e^3 c}{4\pi^2} \left(\dddot{z}^\nu (s) +\dot{z}^\nu (s){\ddot {z}^2(s) } \right) \times \nonumber \\
&& \frac{1}{\dot{z}^0 (s)} \left(\frac{1}{3}\dot{z}_i (s) \dddot {z}_0 (s) + \ddot{z}_i (s) \ddot {z}_0 (s)+\dddot{z}_i (s) \dot {z}_0 (s) \right) \label{ttt}
\end{eqnarray}
%Upto $O(\dddot{z})$ term: 
%\begin{eqnarray}
%&&ec\frac{\dot{z}_\mu (s)} {\dot{z}^0 (s)} F^{\mu \nu}_{rad}(z(s)) (\partial_0 A_i ^{rad}) _{x= z(s)} = \nonumber \\
%&&\frac{2}{3} \frac{e^3 c}{4 \pi ^2} \frac{\ddot{z}_i (s) \ddot {z}_0 (s)}{\dot{z}^0 (s)}  \left(\dddot{z}^\nu (s) +\dot{z}^\nu (s){\ddot {z} }^2 (s) \right) \label{t1} 
%\end{eqnarray}
%%%%%%%%%%%%%%%%%%%%%%%%%%%%%%%%%%%%%%%%%%%
\section*{A3: {Calculation of the term}}
\hspace{0.6in}$\mathbf{
ec \frac{\dot{z}_\mu (s)}{\dot{z}^0 (s)} A_i^{rad}(z(s)) (\partial_0 F^{\mu \nu}_{rad}(x))_{x=z(s)}:}$\\
%%%%%%%%%%%%%%%%%%%%%%%%%%%%%%%%%%%%%%
We have,
\begin{equation}
\partial_0 F^{\mu \nu}_{rad} (x)  =\frac{\partial}{\partial x^0}(\partial^\mu A^\nu (x)-\partial ^\nu A^\mu (x))
\end{equation}
Where,
\begin{eqnarray}
&&\frac{\partial}{\partial x^0}(\partial ^\nu A^\mu (x)) \nonumber \\
&&= e \int ds \, \frac{\partial D(x-z(s))}{\partial x^0}  \frac{d}{ds} \left[\frac{ \dot{z}^\mu (s) (x-z(s))^\nu}{(x-z(s))^{\sigma} \dot{z}_{\sigma}(s)}\right] + \nonumber \\
&& e \int ds\, D(x-z(s))\frac{d}{ds} \frac{\partial}{\partial x^0} \left[\frac{ \dot{z}^\mu (s)(x-z(s))^\nu}{(x-z(s))^{\sigma} \dot{z}_{\sigma} (s)}\right]
 \label{dF0} 
\end{eqnarray}
The first term of the above equation \eqref{dF0} is \\
\begin{eqnarray*}
&&e \int ds \, \frac{\partial D(x-z(s))}{\partial x^0} \frac{d}{ds} \left[\frac{ \dot{z}^\mu (s)(x-z(s))^\nu}{(x-z(s))^{\sigma} \dot{z}_{\sigma}}\right] \nonumber \\
&&= e \int ds \, D(x-z(s)) \frac{d}{d s} \left\lbrace \frac{(x-z(s))^0} {(x-z(s))^{\rho} \dot{z}_{\rho} (s)} \times \right. \nonumber \\
&&\left. \frac{d}{ds} \left[\frac{ \dot{z}^\mu (s)(x-z(s))^\nu}{(x-z(s))^{\sigma} \dot{z}_{\sigma}(s)}\right] \right\rbrace  
\end{eqnarray*}
At $x=z(s^\prime)$ the above equation becomes: 
\begin{eqnarray}
&=&  e \int ds\, D(z(s^\prime)-z(s))\left\lbrace \frac{d}{ds} \left(\frac{(z(s^\prime)-z(s))^0} {(z(s^\prime)-z(s))^{\rho} \dot{z}_{\rho}(s)}\right)  \right. \nonumber\\
&&\left.  \times \frac{d}{ds} \left[\frac{\dot{z}^\mu (s)(z(s^\prime)-z(s))^\nu}{(z(s^\prime)-z(s))^{\sigma} \dot{z}_{\sigma}(s)}\right]\right\rbrace \nonumber   \\
&+& e \int ds \,D(z(s^\prime)-z(s)) \left\lbrace \frac{(z(s^\prime)-z(s))^0} {(z(s^\prime)-z(s))^{\rho} \dot{z}_{\rho}(s)} \right.   \nonumber\\
&&\left.  \times \frac{d^2}{ds^2} \left[\frac{\dot{z}^\mu (s)(z(s^\prime)-z(s))^\nu}{(z(s^\prime)-z(s))^{\sigma} \dot{z}_{\sigma}(s)}\right] \right\rbrace    \label{dF1}
\end{eqnarray}
The first term in equation \eqref{dF1}  can be expressed as
\begin{eqnarray}
&&=\frac{e}{2 \pi} \int du \, \frac{d \delta(u)}{du} \left\lbrace \left[\frac{d}{du} (\dot{z}^0 (s^\prime) \right. \right. \nonumber \\
&& \left. \left. +\frac{u}{2} \ddot{z}^0 (s^\prime) + \frac{u^2}{6}\dddot{z}^0 (s^\prime)....)\right] \right. \nonumber  \\
&&\left. \times \frac{d}{du} \left( \dot{z}^\mu (s^{\prime}) \dot{z}^\nu (s^{\prime})+\frac{u}{2}  \dot{z}^\mu (s^{\prime}) \ddot{z}^\nu (s^{\prime})\right.  \right. \nonumber\\
&&\left. \left. + u \ddot{z}^\mu (s^{\prime}) \dot{z}^\nu (s^{\prime}) + \frac{u^2}{6} \dot{z}^\mu (s^{\prime}) \dddot{z}^\nu (s^{\prime}) \nonumber  \right. \right. \nonumber\\
&&\left. \left. +\frac{u^2}{2} \ddot{z}^\mu (s^{\prime}) \ddot{z}^\nu (s^{\prime})+\frac{u^2}{2} \dddot{z}^\mu (s^{\prime}) \dot{z}^\nu (s^{\prime})  \right. \right. \nonumber\\
&&\left. \left. +\frac{u^3}{6} \ddot{z}^\mu (s^{\prime}) \dddot{z}^\nu (s^{\prime}) +\frac{u^3}{4} \dddot{z}^\mu (s^{\prime}) \ddot{z}^\nu (s^{\prime})... \right)\right\rbrace \nonumber \\
&&=-\frac{e}{2 \pi} \int du \, \delta (u) \left(\frac{d^2 P(u)}{du^2} \frac{d Q(u)}{du} + \frac{dP(u)}{du} \frac{d^2 Q(u)}{du^2}\right) \nonumber \\ \label{dF2}
\end{eqnarray}
where
\begin{eqnarray}
&&P(u)=(\dot{z}^0 (s^\prime)+\frac{u}{2} \ddot{z}^0 (s^\prime) + \frac{u^2}{6}\dddot{z}^0 (s^\prime)....) \nonumber \\
&&Q(u)= \dot{z}^\mu (s^{\prime}) \dot{z}^\nu (s^{\prime})+\frac{u}{2}  \dot{z}^\mu (s^{\prime}) \ddot{z}^\nu (s^{\prime}) \nonumber\\
&&+ u \ddot{z}^\mu (s^{\prime}) \dot{z}^\nu (s^{\prime}) + \frac{u^2}{6} \dot{z}^\mu (s^{\prime}) \dddot{z}^\nu (s^{\prime}) \nonumber  \\
&&+\frac{u^2}{2} \ddot{z}^\mu (s^{\prime}) \ddot{z}^\nu (s^{\prime})+\frac{u^2}{2} \dddot{z}^\mu (s^{\prime}) \dot{z}^\nu (s^{\prime}) \nonumber \\
&& +\frac{u^3}{6} \ddot{z}^\mu (s^{\prime}) \dddot{z}^\nu (s^{\prime}) +\frac{u^3}{4} \dddot{z}^\mu (s^{\prime}) \ddot{z}^\nu (s^{\prime})... 
\end{eqnarray}
The second term in the equation \eqref{dF1} can be expressed as
\begin{eqnarray}
&&=-\frac{e}{2 \pi} \int du \,\delta(u)  \left(\frac{dP(u) }{du} \frac{d^2 Q(u)}{du^2}+P(u)\frac{d^3 Q(u)}{du^3}\right) \nonumber \\
 \label{dF3}
\end{eqnarray}
Now, equation \eqref{dF1} becomes:
\begin{eqnarray}
&&=-\frac{e}{2 \pi} \int du \,\delta(u)  \left\lbrace \frac{d^2 P(u)}{du^2} \frac{d Q(u)}{du}  \right. \nonumber \\
&&\left. +2 \frac{dP(u)}{du} \frac{d^2 Q(u)}{du^2} +P(u)\frac{d^3 Q(u)}{du^3}\right\rbrace  \nonumber \\
&&=-\frac{e}{2 \pi} \left[\frac{\dddot{z}^0 (s)}{3} \left(\frac{\dot{z}^\mu (s) \ddot{z}^\nu (s)}{2}+ \ddot{z}^\mu (s) \dot{z}^\nu (s)\right) \right. \nonumber \\
&& \left. + \ddot{z}^0 (s) \{ \dddot{z}^\nu (s) \frac{\dot{z}^\mu (s)}{3}  (s) + \ddot{z}^\mu (s) \ddot{z}^\nu (s) \right. \nonumber \\
&& \left. + \dddot{z}^\mu (s) \dot{z}^\nu (s) \}  + \dot{z}^0 (s) \{ \ddot{z}^\mu (s) \dddot{z}^\nu (s) \right. \nonumber \\
&&\left.  + \frac{3}{2}\dddot{z}^\mu (s) \ddot{z}^\nu (s) \}  \right] \label{f1}
\end{eqnarray} 
At $x=z(s^\prime)$ the second term in the equation (\ref{dF0}) is
\begin{eqnarray}
&&e \int ds\, D(z(s^\prime)-z(s)) \times \nonumber \\
&&\frac{d}{ds} \frac{\partial}{\partial z^0 (s^\prime)} \left[\frac{ \dot{z}^\mu (s)(z(s^\prime)-z(s))^\nu}{(z(s^\prime)-z(s))^{\sigma} \dot{z}_{\sigma} (s)}\right] \nonumber \\
&&= -\frac{e}{2 \pi} \int du \,  \frac{1}{\dot{z}^0 (s^\prime)}\delta (u) \times \nonumber \\
&&\frac{d^2}{du^2}   \left( \dot{z}^\mu (s^{\prime}) \ddot{z}^\nu (s^\prime) +\ddot{z}^\mu (s^{\prime}) \dot{z}^\nu  (s^{\prime}) +\frac{u}{2}  \dot{z}^\mu (s^{\prime}) \dddot{z}^\nu (s^{\prime})\nonumber \right.\\
&&\left. +\frac{3u}{2}  \ddot{z}^\mu (s^{\prime}) \ddot{z}^\nu (s^{\prime})+ u \dddot{z}^\mu (s^{\prime}) \dot{z}^\nu (s^{\prime})  \nonumber \right. \\
&& \left. + \frac{2u^2}{3} \ddot{z}^\mu (s^{\prime}) \dddot{z}^\nu (s^{\prime})+ u^2 \dddot{z}^\mu (s^{\prime}) \ddot{z}^\nu (s^{\prime}) \right) \nonumber \\
&&=-\frac{e}{2 \pi} \frac{1}{\dot{z}^0 (s)} \left(\frac{4}{3}\dddot{z}^\nu (s)\ddot{z}^\nu (s) + 2 \dddot{z}^\mu (s) \ddot{z}^\nu (s) \right) 
 \label{f2} 
\end{eqnarray}
By substituting equations (\ref{f1}) and (\ref{f2}) in the equation (\ref{dF0}), we have,
\begin{eqnarray}
&&\frac{\partial}{\partial x^0}(\partial ^\nu A^\mu (x))_{x=z(s)} \nonumber \\
&&=-\frac{e}{2 \pi} \left[\frac{1}{3} \dddot{z}^0 (s) \left\lbrace \frac{1}{2} \dot{z}^\mu (s) \ddot{z}^\nu (s)+ \ddot{z}^\mu (s) \dot{z}^\nu (s)\right\rbrace  \right.\nonumber \\
&&\left. + \ddot{z}^0 (s) \left\lbrace \frac{1}{3} \dddot{z}^\nu (s) \dot{z}^\mu (s) \nonumber + \ddot{z}^\mu (s) \ddot{z}^\nu (s) +\dddot{z}^\mu (s) \dot{z}^\nu (s)\right\rbrace \right.   \nonumber \\
&& \left. + \dot{z}^0 (s) \left\lbrace \ddot{z}^\mu (s) \dddot{z}^\nu (s)+ \frac{3}{2}\dddot{z}^\mu (s) \ddot{z}^\nu (s) \right\rbrace  \right. \nonumber  \\
&&\left. + \frac{1}{\dot{z}^0 (s)} \left\lbrace \frac{4}{3}\dddot{z}^\nu (s)\ddot{z}^\mu (s)+ 2 \dddot{z}^\mu (s) \ddot{z}^\nu (s) \right\rbrace  \right]
\end{eqnarray}
Now:
\begin{eqnarray}
&&(\partial _0 F^{\mu \nu} (x))_{x=z(s)} \nonumber \\
&& =\frac{e}{2 \pi} \left[\frac{\dddot{z}^0 (s)}{6} \left(\ddot{z}^\mu (s) \dot{z}^\nu (s)-\ddot{z}^\nu (s) \dot{z}^\mu (s)\right) \right.  \nonumber \\
&&\left.  + \frac{2}{3}  \ddot{z}^0 (s) \left( \dddot{z}^\mu (s) \dot{z}^\nu (s) -\dddot{z}^\nu (s) \dot{z}^\mu (s) \right) \right.   \nonumber \\
&&\left. +   \left( \dddot{z}^\mu (s) \ddot{z}^\nu (s) -\dddot{z}^\nu (s) \ddot{z}^\mu (s) \right) \left( \frac{1}{2} \dot{z}^0 (s)+ \frac{2}{3} \frac{1}{\dot{z}^0 (s)} \right)\right]\nonumber
\end{eqnarray}
Therefore
\begin{eqnarray}
&&ec \frac{\dot{z}_\mu (s)}{\dot{z}^0 (s)} A_i^{rad}(z(s)) (\partial_0 F^{\mu \nu}_{rad}(x))_{x=z(s)} \nonumber \\
&&= \frac{e^3 c}{4 \pi ^2} \frac{\ddot{z}_i (s)}{\dot{z}^0 (s)} \left[\frac{ \dddot{z}^0 (s) \ddot{z}^\nu (s)}{6} \right. \nonumber \\
&& \left. + \frac{2}{3}\ddot{z}^0 (s) \left(  \dddot{z}^\nu (s)  + \ddot{z} ^2 (s) \dot{z}^\nu (s) \right) \right. \nonumber \\
&&{} \left. +  \ddot{z} ^2 (s) \ddot{z}^\nu (s)  \left(\frac{1}{2} \dot{z}^0 (s)+ \frac{2}{3} \frac{1}{\dot{z}^0 (s)} \right)\right] \label{fdf}
\end{eqnarray}
%%%%%%%%%%%%%%%%%%%%%%%%%%%%%%%%%%%%%%%%%%%%%%%%%%%%%%
\section*{A4: Calculation of the term}
\hspace{0.5in}
$\mathbf{ec \frac{A^{rad}_i(z(s))}{\dot{z}^0 (s)} {F^{\mu \nu}_{rad}(z(s))} \frac{\partial}{\partial s} \left(\frac{\dot{z}_\mu (s)}{\dot{z}^0 (s)}\right):}$\\
We have,
\begin{eqnarray*}
&&\frac{A^{rad}_i( z(s))}{\dot{z}^0 (s)} {F^{\mu \nu}_{rad}(z(s))} \frac{\partial}{\partial s} \left[\frac{\dot{z}_\mu (s)}{\dot{z}^0 (s)}\right]\\
&&=\frac{\ddot{z}_\mu (s)}{(\dot{z}^0 (s))^2 } F^{\mu \nu}_{rad}(z(s)) A_i^{rad} (z(s)) 
- \frac{\dot{z}_{\mu} \ddot{z}^0}{(\dot{z}^0)^3}  F^{\mu \nu}_{rad}(z) A_i^{rad} (z) 
\end{eqnarray*}
Now,
\begin{eqnarray}
&& \frac{\ddot{z}_\mu (s)}{(\dot{z}^0 (s))^2  } F^{\mu \nu}_{rad}(z(s)) A_i^{rad}(z(s)) \nonumber \\
&&= - \frac{2}{3} \frac{e^2}{4 {\pi}^2} \frac{\ddot{z}_i (s)}{(\dot{z}^0 (s))^2}\left(\dot{z}^\mu (s) \ddot{z}_\mu (s) \dddot{z}^\nu (s)-\dot{z}^\nu (s) \ddot{z}_\mu (s) \dddot{z}^\mu (s) \right)\nonumber \\
&&{}=  \frac{2}{3} \frac{e^2}{4 {\pi}^2} \frac{\ddot{z}_i (s)}{(\dot{z}^0 (s))^2} \dot{z}^\nu (s) \ddot{z}_\mu (s) 
\dddot{z}^\mu (s) \label{t4}
\end{eqnarray}
%$\dot{z}^\mu \ddot{z}_\mu =0$ and $\ddot{z}_\mu \dddot{z}^\mu = -\frac{1}{3}\dot{z}_\mu \ddddot{z}^\mu$\\
%We are neglecting $\ddddot{z}$ and other higher derivatives.Therefore there is no contribution to radiation reaction from the term
%$\frac{\ddot{z}_\mu (s)}{(\dot{z}^0 (s))^2 } F^{\mu \nu}_{rad}(z(s)) A_i^{rad} (z(s))$.\\
And,
\begin{eqnarray}
&& \frac{\ddot{z}^0 (s) \dot{z}_\mu (s)}{(\dot{z}^0 (s))^3  } F^{\mu \nu}_{rad}(z(s)) A_i^{rad} (z(s)) \nonumber \\
&&=\frac{2}{3} \frac{e^2}{4{\pi}^2} \frac{\ddot{z}_i (s) \ddot{z}^0 (s)}{(\dot{z}^0 (s))^3 } \left(\dddot{z}^\nu (s) + 
\ddot{z}^2 (s) \dot{z}^\nu (s) \right) \label{t44}
\end{eqnarray}
Therefore
\begin{eqnarray}
&&ec \frac{A^{rad}_i(z(s))}{\dot{z}^0 (s)} {F^{\mu \nu}_{rad}(z(s))} \frac{\partial}{\partial s} ( \frac{\dot{z}_\mu (s)}{\dot{z}^0 (s)}) \nonumber \\
&&=\frac{2}{3} \frac{e^3 c}{4 {\pi}^2} \frac{\ddot{z}_i (s)}{(\dot{z}^0 (s))^2} \dot{z}^\nu (s) \ddot{z}_\mu (s) \dddot{z}^\mu (s) \nonumber \\
&&-\frac{2}{3} \frac{e^3 c}{4{\pi}^2} \frac{\ddot{z}_i (s) \ddot{z}^0 (s)}{(\dot{z}^0 (s))^3} \left(\dddot{z}^\nu (s) + 
\ddot{z}^2 (s) \dot{z}^\nu (s) \right)
\end{eqnarray}
From equations \eqref{ttt} and \eqref{fdf}
\begin{eqnarray}
&& \int \frac{\partial}{\partial x^0} (A_i ^{rad}  F^{\mu \nu}_{rad}) J_\mu d^3 x  \nonumber \\
&&=\frac{2}{3} \frac{e^3 c}{4 \pi ^2} \left(\dddot{z}^\nu (s) +\dot{z}^\nu (s){\ddot {z}^2(s) } \right) \times \nonumber \\
&& \frac{1}{\dot{z}^0 (s)} \left(\frac{1}{3}\dot{z}_i (s) \dddot {z}_0 (s) + \ddot{z}_i (s) \ddot {z}_0 (s)+\dddot{z}_i (s) 
\dot {z}_0 (s) \right) \nonumber \\
&& + \frac{e^3 c}{4 \pi ^2} \frac{\ddot{z}_i (s)}{\dot{z}^0 (s)} \left[\frac{ \dddot{z}^0 (s) \ddot{z}^\nu (s)}{6} + 
\frac{2}{3}\ddot{z}^0 (s) \left(  \dddot{z}^\nu (s)  + \ddot{z} ^2 (s) \dot{z}^\nu (s) \right) \right. \nonumber \\
&& \left. +  \ddot{z} ^2 (s) \ddot{z}^\nu (s)  \left(\frac{1}{2} \dot{z}^0 (s)+ \frac{2}{3} \frac{1}{\dot{z}^0 (s)} \right)\right] \label{fin1}
\end{eqnarray}
From equation \eqref{t4} and \eqref{t44}, we have
\begin{eqnarray}
&&\int \frac{\partial J_\mu}{\partial x^0} (A_i ^{rad} F^{\mu \nu}_{rad})  d^3 x \nonumber \\
&&=\frac{2}{3} \frac{e^3 c}{4 {\pi}^2} \frac{\ddot{z}_i (s)}{(\dot{z}^0 (s))^2} \dot{z}^\nu (s) \ddot{z}_\mu (s) \dddot{z}^\mu (s) \nonumber \\
&&-\frac{2}{3} \frac{e^3 c}{4{\pi}^2} \frac{\ddot{z}_i (s) \ddot{z}^0 (s)}{(\dot{z}^0 (s))^3} \left(\dddot{z}^\nu (s) +
\ddot{z}^2 (s) \dot{z}^\nu (s) \right)
 \label{fin2}
\end{eqnarray}
Radiation reaction becomes:
\begin{eqnarray}
&&G^{\nu}_{NC}= -\frac{2}{3} \frac{e^2}{4 \pi } \frac{1}{\dot{z}^0 (s)} \left(\dddot{z}^\nu (s) + \dot{z}^\nu (s) \ddot {z}^2 (s)  \right)\nonumber\\
&&+{\theta}^{0i} \frac{e^3}{16{\pi}^2}  \nonumber \\
&&\left\lbrace \frac{\ddot{z}^i (s)}{\dot{z}^0 (s)}  \left[\frac{2}{3}\ddot{z}^0 (s) \left(\dddot{z}^\nu (s) + \ddot{z}^2 (s) \dot{z}^\nu (s) \right)\left(\frac{1}{(\dot{z}^0 (s))^2}-2\right) \nonumber \right. \right. \\
&& \left. \left. -  \frac{1}{6} \dddot{z}^0 (s) \ddot{z}^\nu (s) -  \ddot{z} ^2 (s) \ddot{z}^\nu (s)  \left(\frac{1}{2} \dot{z}^0 (s)+ \frac{2}{3} \frac{1}{\dot{z}^0 (s)} \right) \nonumber \right. \right. \\
&& \left. \left. - \frac{\dot{z}^\nu (s) \ddot{z}_\mu (s) \dddot{z}^\mu (s)}{\dot{z}^0 (s)}  \right]  - \frac{2}{3} \left(\dddot{z}^\nu (s) +\dot{z}^\nu (s){\ddot {z}^2(s) } \right) \times \right.  \nonumber \\
&& \left. \frac{1}{\dot{z}^0 (s)} \left(\frac{1}{3}\dot{z}^i (s) \dddot {z}_0 (s) +\dddot{z}^i (s) \dot {z}_0 (s) \right)\right\rbrace 
\end{eqnarray}
Where,
\begin{eqnarray}
&&\frac{\ddot{z}^i (s)\ddot{z}^0 (s)}{\dot{z}^0 (s)} \left(\dddot{z}^j (s) + \dot{z}^j (s) \ddot {z}^2 (s)  \right)\left(\frac{1}{(\dot{z}^0 (s))^2}-2\right) \nonumber   \\
&&=- \frac{1}{\gamma} \frac{\gamma ^2}{c^2} \left( \dot{v}_i + \frac{\gamma ^2}{c^2} v_i (\vec{v} . \dot{\vec{v}}) \right) \frac{\gamma ^4}{c^3}  (\vec{v} . \dot{\vec{v}}) \frac{\gamma ^3}{c^3} \times \nonumber \\
&&\left(\ddot{v}_j +\frac{3 \gamma ^2}{c^2} \dot{v}_j (\vec{v}.\dot{\vec{v}}) + \frac{\gamma ^2}{c^2} v_j (\vec{v}. \ddot{\vec{v}})+\frac{3 \gamma ^4}{c^4} (\vec{v}.\dot{\vec{v}})^2 v_j \right) \left(1+\frac{v^2}{c^2}\right) \nonumber \\
&&=-\frac{\gamma ^8}{c^{8}}   \dot{v}_i (\vec{v}.\dot{\vec{v}}) \ddot{v}_j - \frac{\gamma ^8}{c^{10}} v^2  \dot{v}_i (\vec{v}.\dot{\vec{v}}) \ddot{v}_j 
\end{eqnarray}
with,
\begin{eqnarray}
&&\frac{\ddot{z}^i (s)}{\dot{z}^0 (s)} \ddot{z}^j \dddot{z}^0 (s)=\frac{1}{\gamma} \frac{\gamma ^8}{c^8} \left( \dot{v}_i + \frac{\gamma ^2}{c^2} v_i (\vec{v} . \dot{\vec{v}}) \right)  \times \nonumber \\
&& \left( \dot{v}_j + \frac{\gamma ^2}{c^2} v_j (\vec{v} . \dot{\vec{v}}) \right) \left((\vec{v}. \ddot{\vec{v}})+ \dot{v} ^2 + 4 \frac{\gamma ^2}{c^2} (\vec{v}.\dot{\vec{v}})^2 \right) \nonumber\\
&&=\frac{\gamma ^8}{c^8}  \dot{v}_i \left((\vec{v}. \ddot{\vec{v}})+ \dot{v} ^2 \right)\dot{v}_j 
\end{eqnarray}
And,
\begin{eqnarray}
&&\frac{\ddot{z}^i (s)}{\dot{z}^0 (s)} \ddot{z}^2 (s) \ddot{z}^j (s) \left(\frac{z^0 (s)}{2}+ \frac{2}{3 \dot{z}^0 (s)}\right) \nonumber \\
%&&=\ddot{z}^i (s)\ddot{z}^j (s) \ddot{z}^2 (s)  \left(\frac{1}{2}+ \frac{2}{3 (\dot{z}^0 (s))^2}\right)\nonumber \\
&&= \frac{\gamma ^2}{c^2} \left( \dot{v}_i + \frac{\gamma ^2}{c^2} v_i (\vec{v} . \dot{\vec{v}}) \right)\frac{\gamma ^2}{c^2} \left( \dot{v}_j + \frac{\gamma ^2}{c^2} v_j (\vec{v} . \dot{\vec{v}}) \right) \times \nonumber \\
&& (-\frac{\gamma ^4}{c^4}) \left(\dot{v}^2 + 2 \frac{\gamma ^2}{c^2} (\vec{v} . \dot{\vec{v}})^2 +  \frac{\gamma ^4}{c^4} v^2 (\vec{v} . \dot{\vec{v}})^2 - \frac{\gamma ^4}{c^2}  (\vec{v} . \dot{\vec{v}})^2 \right) \times  \nonumber \\
&& \left(\frac{1}{2}+ \frac{2}{3}(1-\frac{v^2}{c^2})\right) \nonumber \\
&&=-\frac{7 \gamma ^8}{6 c^8} \dot{v}_i \dot{v} ^2 \dot{v}_j + \frac{2 \gamma ^8}{3 c^{10}} \dot{v}_i \dot{v} ^2 \dot{v}_j v^2 
\end{eqnarray}
Here,
\begin{eqnarray}
\frac{\ddot{z}^i}{(\dot{z}^0)^2} \dot{z}^j \ddot{z}_\mu \dddot{z}^\mu 
&=&-\frac{\gamma^6}{c^6} \dot{v}_i v_j \dot{\vec{v}}.\ddot{\vec{v}}- 3 \frac{\gamma^8}{c^8} \dot{v}_i v_j \dot{v}^2 \vec{v}.\dot{\vec{v}}\nonumber \\
&-&\frac{\gamma ^8}{c^8} \dot{v}_i v^j \vec{v}.\dot{\vec{v}} \left(\dot{v}^2 + \vec{v}.\ddot{\vec{v}}\right)\nonumber \\
&-&\frac{\gamma ^8}{c^8} \dot{v}^i v^j  \left(\vec{v}.\dot{\vec{v}}\right)  \left(\vec{v}.\ddot{\vec{v}}\right)\nonumber \\
&-&\frac{\gamma ^8}{c^{10}} v^i v^j  \left(\vec{v}.\dot{\vec{v}}\right)  \left(\dot{\vec{v}}.\ddot{\vec{v}}\right)
\end{eqnarray}
And,
\begin{eqnarray}
&&\frac{1}{\dot{z}^0} \left(\dddot{z}^i \dot{z}^0 + \frac{1}{3}\dot{z}^i \dddot{z}^0 \right) \left(\dddot{z}^j + \dot{z}^j \ddot{z}^2 \right) \nonumber \\
&&= \frac{\gamma^6}{c^6} \ddot{v}_i \ddot{v}_j + 3 \frac{\gamma^8}{c^8}\ddot{v}_i \dot{v}_j \left(\vec{v}.\dot{\vec{v}}\right)\nonumber \\
&&+ \frac{\gamma^8}{c^8}\ddot{v}_i v_j \left(\vec{v}.\ddot{\vec{v}}\right)+ 3 \frac{\gamma^8}{c^8}\dot{v}_i \ddot{v}_j \left(\vec{v}.\dot{\vec{v}}\right) \nonumber \\
&&+ \frac{4}{3} \frac{\gamma^8}{c^8} v_i \ddot{v}_j \left(\dot{v}^2 + \vec{v}.\ddot{\vec{v}}\right)
\end{eqnarray}

\end{document}